\documentclass[a4paper,12pt]{article}
\usepackage[utf8]{inputenc}
\usepackage{graphicx} 
\usepackage{amsmath}
\usepackage{textcomp}
\usepackage{amsfonts}
\usepackage{multicol}
\usepackage{cite}

\normalbaselines
\title{Supersymmetric partners of the harmonic oscillator with an infinite potential barrier}

\author{DJ Fern\'andez and VS Morales-Salgado
 \\ {\sl Departamento de F\'{\i}sica, Cinvestav  A.P. 14-740,}
\\ {\sl 07000 M\'exico D.F.  Mexico}}

\begin{document}

\maketitle

\begin{abstract}
Supersymmetry transformations of first and second order are used to generate Hamiltonians with known spectra 
departing from the harmonic oscillator with an infinite potential barrier. It is studied also the way in which 
the eigenfunctions of the initial Hamiltonian are transformed. The first and certain second order supersymmetric 
partners of the initial Hamiltonian possess third-order differential ladder operators. Since systems with this 
kind of operators are linked with the Painlev\'e IV equation, several solutions of this non-linear 
second-order differential equation will be simply found.
\end{abstract}

\section{Introduction}
Supersymmetric Quantum Mechanics (SUSY QM) has proven to be an exceptional technology for generating quantum mechanical 
potentials with known spectra \cite{mi84,cks95,jr98,mnn98,qv99,mr04,cfnn04,cf08,ma09,fe10,q11,bf11a,ma12,ai12,ggm13}. 
In this method the spectrum of an initial Hamiltonian is modified, by creating or deleting levels, in order to implement 
the so-called spectral design \cite{cf08,fe10,ai12}. An important fact is that the intertwining operator technique and 
the factorization method are procedures which are equivalent to SUSY QM \cite{fe10}.

It is well known that the harmonic oscillator Hamiltonian has an equidistant energy spectrum, which is 
due to its intrinsic algebraic structure known as the Heisenberg-Weyl algebra. On the other hand, the 
polynomial Heisenberg algebras are deformations of the oscillator algebra, where the differential ladder 
operators are of order $m+1$ and the commutator between them is a polynomial of order $m$ in the 
Hamiltonian. Due to this algebraic structure, the Hamiltonian spectrum turns out to be the juxtaposition 
of several equidistant energy ladders. 

Let us note that systems described by second order polynomial Heisenberg algebras (for $m=2$) 
are connected to the Painlev\'e IV (PIV) equation \cite{srk97,ad94,vs93,ain95,acin00,bch95,sh92,dek94,ekk94,mn08}. 
Conversely, if a system characterized by third order differential ladder operators and their extremal states are found, 
thus one can find solutions to the PIV equation in a simple way. 

The supersymmetric partners of the harmonic oscillator Hamiltonian provide explicit realizations of the 
polynomial Heisenberg algebras \cite{fh99,mn08,cfnn04,fnn04}. In particular, for first order SUSY 
the involved supercharges are linear in the momentum and such Hamiltonians have 
third order ladder operators \cite{mi84}, i.e., they fulfill the commutation relations associated 
to a second order polynomial Heisenberg algebra, thus they will lead to solutions to the 
PIV equation \cite{cfnn04}. 

On the other hand, Hamiltonians obtained from the harmonic oscillator through second order SUSY 
with supercharges which are quadratic in the momentum
possess, in general, fifth order differential ladder operators. However, it is possible to identify
a subfamily of these Hamiltonians which, in addition to have these fifth order operators, possesses also 
third order ones and, consequently, lead to new solutions of the PIV equation \cite{bf11a}. The same 
property can be found for a subset of $k$-th order SUSY partner Hamiltonians of the oscillator, which have 
the two kinds of differential ladder operators, those of order $2k+1$ and third order ones, 
the last leading also to solutions to the PIV equation.

Using the previously mentioned technique, plenty of non-singular explicit solutions to the PIV equation, 
either real or complex, have been derived \cite{bf11a,bf11b,be12}
(for other methods to generate solutions to the PIV equation the reader can seek \cite{bch95,gls02}).
It would be important to address a systematic analysis of the 
corresponding singular solutions. In this paper we will start this study by allowing the existence of one fixed 
singularity. Our treatment will be based on the harmonic oscillator with an infinite potential barrier \cite{mnn98}, 
which for simplicity will be placed at the origin. We will be mainly interested in transformations which reproduce 
again the singularity present in the initial potential, i.e., PIV solutions having singularities for $x=0$. 
We shall describe also the induced spectral modifications and the second order polynomial Heisenberg algebra 
characterizing the new Hamiltonians, which will naturally lead to new solutions to the PIV equation.

In order to achieve our goals, in Section 2 we will review briefly the Supersymmetric Quantum Mechanics and the way 
in which Hamiltonians which are intertwined with the harmonic oscillator realize the second order polynomial 
Heisenberg algebras, connecting them later with the PIV equation and some of its solutions. In Section 3 we 
will study the harmonic oscillator with and infinite potential barrier at the origin, and we will apply to it the 
first and second order SUSY techniques. In Section 4 we will obtain several solutions to the PIV equation, 
either non-singular or with a singularity at $x=0$, by using the extremal states of the SUSY partners of the 
harmonic oscillator with an infinite potential barrier. Finally, in Section 5 we will emphasize the original results 
obtained in this paper as well as our conclusions.

\section{Supersymmetric Quantum Mechanics}\label{SQM}

Supersymmetric Quantum Mechanics describes systems characterized by a supersymmetric Hamiltonian $H_{\footnotesize \mbox{ss}}$ 
and two supercharges $Q_1$, $Q_2$, all of them hermitian operators satisfying the following supersymmetry 
algebra with two generators \mbox{\cite{wi81,wi82,ais93,aicd95,bs97,fgn98,fhm98,sa99,mnr00,crf01,ast01,in04,ff05}}:
\begin{equation}\label{0.3.3}
[H_{\scriptsize \mbox{ss}},Q_i]=0, \qquad \{Q_i,Q_j\}=\delta_{ij}H_{\scriptsize \mbox{ss}}, \quad i,j=1,2,
\end{equation}
where $[F,G]=FG-GF$ and $\{F,G\}=FG+GF$ are the commutator and anticommutator of the operators $F$ and $G$ 
respectively. 

The simplest realizations of such an algebra arise from the intertwining operator technique as follows \cite{ff05}. 
Let us suppose that a pair of Hamiltonians 
\begin{equation}\label{0.1.1}
H=-\frac{1}{2}\frac{\rm d^2}{\rm d x^2}+V(x),\quad \tilde{H}=-\frac{1}{2}\frac{\rm d^2}{\rm d x^2}+\tilde{V}(x),
\end{equation}
where $V(x)$ and $\tilde{V}(x)$ are real potentials, obey the intertwining relations
\begin{equation}\label{0.1.2}
\tilde{H}A^+=A^+H \quad \Leftrightarrow \quad HA=A\tilde{H},
\end{equation}
with $A^+$ and $A$ being differential intertwining operators of order $k$ 
(we are using units such that $\hbar=m=1$). These operators satisfy
\begin{equation}\label{0.2.2}
AA^+= \prod_{i=1}^{k}(H-\epsilon_i), \qquad A^+A= \prod_{i=1}^{k}(\tilde{H}-\epsilon_i), \qquad \epsilon_i \in \mathbb{R}.
\end{equation}

The requirement $\epsilon_i \in \mathbb{R}$, $i=1,...,k$ is taken mainly by two reasons: 
(i) $\tilde{V}(x)$ should be real; (ii) also we will look for real solutions to the PIV equation.
In fact, if we just would require that $\tilde{V}(x)$ be real, without worrying about the PIV solution,
(\ref{0.2.2}) could include pairs of complex conjugate factorization energies $\epsilon_j$, $\bar{\epsilon}_j$,
leading to a real SUSY partner potential \cite{aicd95,ff05,ai12}. 
Moreover, this transformation can be decomposed into first and second-order ones, 
the second-order transformations involving $\epsilon_j$ and $\bar{\epsilon}_j$ in an irreducible way, i.e.,
they also can be produced by two first-order transformations but the intermediate potential would be complex \cite{aicd95}.
For the purposes of this paper it is enough to assume that $\epsilon_i \in \mathbb{R}$, $i=1,..k$, 
and the same will be done further on for any other factorization of this kind.

In order to realize the standard supersymmetry algebra of (\ref{0.3.3}) let us choose
\begin{equation}\label{0.3.1}
Q_1=\frac{Q^++Q^-}{\sqrt{2}}, \quad Q_2=\frac{Q^+-Q^-}{\rm i\sqrt{2}}, \quad Q^+=\begin{pmatrix} 0&A^+\\ 0&0 \end{pmatrix}, 
\quad Q^-=\begin{pmatrix} 0&0\\ A&0 \end{pmatrix},
\end{equation}
so that
\begin{equation}\label{0.3.2}
H_{\scriptsize \mbox{ss}}=\{Q^-,Q^+\}=\begin{pmatrix}A^+A & 0  \\ 0 & AA^+ \end{pmatrix} .
\end{equation}
Since $A^+$ and $A$ are the previous $k$-th order differential intertwining operators, this representation 
is known as $k$-SUSY QM. Hence, the supersymmetric Hamiltonian $H_{\scriptsize \mbox{ss}}$ turns out to be
\begin{equation}\label{0.3.4}
 H_{\scriptsize \mbox{ss}}=(H_{\scriptsize \mbox{d}}-\epsilon_1)\dots(H_{\scriptsize \mbox{d}}-\epsilon_k),
\end{equation}
where
\begin{equation}\label{0.3.5}
 H_{\scriptsize \mbox{d}}-\epsilon_i=\begin{pmatrix}\tilde{H}-\epsilon_i & 0  \\ 0 & H-\epsilon_i\end{pmatrix}, \qquad i=1,...,k.
\end{equation}
 
\subsection{1-SUSY}\label{eprimero}

Let the operators $A^+$ and $A$ be of first order \cite{wi81,wi82,cks95,mr04,ff05,ai12}, i.e.,
\begin{equation}\label{0.1.3}
A^+=\frac{1}{\sqrt{2}}\left[-\frac{\rm d}{\rm d x}+\alpha(x)\right], \qquad  A=\frac{1}{\sqrt{2}}\left[\frac{\rm d}{\rm d x}+\alpha(x)\right],
\end{equation}
where $\alpha(x)$ is a real function of $x$. By plugging these expressions in the intertwining relations (\ref{0.1.2})
one gets that $\alpha$ must satisfy: 
\begin{equation}\label{riccati}
\alpha'+\alpha^2=2\left(V-\epsilon\right).
\end{equation}
Moreover, the substitution $\alpha=\left[\ln(u)\right]'=u'/u$ transforms this Riccati equation for $\alpha$ into 
a stationary Schr\"odinger one for $u$,
\begin{equation}\label{MQS4}
 -\frac{1}{2}u''+Vu=Hu=\epsilon u,
\end{equation}
where $\epsilon$ is a real constant called {\it factorization energy}. Besides (\ref{riccati}), it is obtained
the following expression for the potential $\tilde{V}(x)$:
\begin{equation}\label{0.1.8a}
\tilde{V} = V- \alpha' = V - \left[\ln(u)\right]''.
\end{equation}
Hence, if we choose a nodeless {\it seed solution} $u$ of the stationary Schr\"odinger equation (also called 
{\it transformation function}) associated to a given factorization energy $\epsilon$, then the intertwining operators 
$A^+, \ A,$ and the new Hamiltonian $\tilde{H}$ become completely determined. Moreover, departing from the normalized 
eigenfunctions $\psi_n(x)$ of $H$ associated to the eigenvalues $E_n$, the corresponding ones $\phi_n(x)$ of $\tilde{H}$ 
are typically found through 
\begin{equation}\label{0.2.8b}
\phi_n(x)=\frac{A^+\psi_n(x)}{\sqrt{E_n-\epsilon}}.
\end{equation}
An additional eigenfunction $\phi_\epsilon(x)$ of $\tilde{H}$, associated to the eigenvalue $\epsilon$, could exist, which 
obeys
\begin{equation}
 A\phi_\epsilon(x) = 0 \quad \Rightarrow \quad \phi_\epsilon(x) \propto \exp\left[-\int \alpha(x) \rm d x \right] \propto 1/u(x).
\end{equation}
Since $\tilde{H} \phi_\epsilon(x) = \epsilon \phi_\epsilon(x)$, then if $\phi_\epsilon(x)$ satisfies the given boundary 
conditions it turns out that $\epsilon$ must be incorporated to the set of eigenvalues of $\tilde{H}$.

\subsection{2-SUSY}\label{esegundo}

Let us suppose now that the intertwining operators $A^+$ and $A$ are of second order \cite{ais93,aicd95,bs97,fgn98,fhm98,sa99,mnr00,crf01,ff05}, i.e.,
\begin{equation}\label{0.1.9} 
A^+ = \frac{1}{2}\left[\frac{\rm d^2}{\rm d x^2}-\eta(x)\frac{\rm d}{\rm d x}+\gamma(x)\right], \qquad
A = \frac{1}{2}\left[\frac{\rm d ^2}{\rm d x^2}+\eta(x)\frac{\rm d}{\rm d x}+\eta'(x)+\gamma(x)\right].
\end{equation}
A similar treatment as for 1-SUSY leads now to the following non-linear second-order differential
equation for $\eta$:
\begin{equation}
 \frac{\eta\eta''}2 - \frac{(\eta')^2}4 + \eta^2\left(\eta' + \frac{\eta^2}4 - 2V + \epsilon_1 + \epsilon_2\right) + (\epsilon_1 - \epsilon_2)^2 = 0 .
\end{equation}
Its solutions, in terms of either two solutions $\alpha_{1,2}$ of the Riccati equation associated to 
$\epsilon_{1,2}$ or the corresponding Schr\"odinger ones $u_{1,2}$ for $\epsilon_1 \neq \epsilon_2$,
read:
\begin{equation}
 \eta = - \frac{2(\epsilon_1 - \epsilon_2)}{\alpha_1 - \alpha_2} = \left[\ln W(u_1,u_2)\right]',
\end{equation}
where $W(u_1,u_2) = u_1 u_2' - u_1' u_2$ is the Wronskian of $u_1$ and $u_2$. Two additional expressions arise from the
intertwining relations (\ref{0.1.2}):
\begin{eqnarray}
  \gamma = \frac{\eta'}2 + \frac{\eta^2}2 - 2V + \epsilon_1 + \epsilon_2, \\
  \tilde V = V - \eta' = V - \left[\ln W(u_1,u_2)\right]''. \label{0.1.28}
\end{eqnarray}
Hence, if we choose now two seed solutions $u_{1,2}$ of the stationary Schr\"odinger equation associated to 
$\epsilon_{1,2}$ such that $W(u_1,u_2)$ is nodeless inside the domain of $V(x)$, it turns out that $ A^+, \ A,$ and 
$\tilde H$ become once again completely determined. Moreover, the eigenfunctions $\phi_n(x)$ of $\tilde H$ associated to 
the eigenvalues $E_n$ become obtained typically from those $\psi_n(x)$ of $H$ through the standard expression:
\begin{equation}\label{0.2.16}
 \phi_n(x)=\frac{A^+\psi_n(x)}{\sqrt{(E_n-\epsilon_1)(E_n-\epsilon_2)}}.
\end{equation}
Two extra eigenfunctions $\phi_{\epsilon_{1,2}}(x)$ of $\tilde{H}$, associated to the eigenvalues $\epsilon_{1,2}$, 
could exist, which obey \cite{ff05}
\begin{equation}\label{twomissing}
 A\phi_{\epsilon_{1,2}}(x) = 0 \qquad \Rightarrow \qquad \phi_{\epsilon_1}(x) \propto \frac{u_2}{W(u_1,u_2)}, \qquad \phi_{\epsilon_2}(x) \propto \frac{u_1}{W(u_1,u_2)}.
\end{equation}
Since $\tilde{H} \phi_{\epsilon_{1,2}}(x) = \epsilon_{1,2} \phi_{\epsilon_{1,2}}(x)$, then if the two 
$\phi_{\epsilon_{1,2}}(x)$ satisfy the given boundary conditions it turns out that $\epsilon_{1,2}$ must be included 
in the spectrum of $\tilde{H}$.

\subsection{Polynomial Heisenberg Algebras}

The polynomial Heisenberg algebras are deformations of the harmonic oscillator algebra, which are characterized by two 
standard commutation relations
\begin{equation}\label{0.4.1}
 [{\mathbb H},L^\pm]=\pm L^\pm,
\end{equation}
plus one defining the deformation
\begin{equation}\label{0.4.2}
 [L^-,L^+]\equiv N({\mathbb H} + 1)-N({\mathbb H})=P_m({\mathbb H}),
\end{equation}
where the analogue to the number operator, $N({\mathbb H})\equiv L^+L^-$, is a polynomial of degree 
$m+1$ in the Hamiltonian ${\mathbb H}$ so that $P_m({\mathbb H})$ becomes of degree $m$ \cite{fnn04}. 
Note that $N({\mathbb H})$ admits the following factorization: 
\begin{equation}\label{0.4.3}
  N({\mathbb H})=\prod_{i=1}^{m+1}({\mathbb H} - \varepsilon_i).
\end{equation}

Let us realize now the polynomial Heisenberg algebras of (\ref{0.4.1}-\ref{0.4.3}) by using the intertwining 
operator technique of section 2. In order to do that, let us express first the commutation relation which involves 
${\mathbb H}$ and $L^+$ in the standard intertwining form:
\begin{equation}
({\mathbb H} - 1) L^+ = L^+ {\mathbb H}.
\end{equation}
By comparing this with (\ref{0.1.2}) it is natural to make $H = {\mathbb H}$, $\tilde H = {\mathbb H} - 1$, 
$A^+ = L^+, \ A = L^-$, $k=m+1$ and $\epsilon_i = \varepsilon_i - 1$. Thus, (\ref{0.2.2}) automatically leads to 
the commutation relation of (\ref{0.4.2}). In this way it is obtained a realization of the poynomial Heisenberg 
algebras of (\ref{0.4.1}-\ref{0.4.3}) in terms of differential operators of finite order. There is, however, 
an important difference that must be stressed: while in the first part of section 2 it was assumed that $H$ is known in order to 
generate $\tilde H$, now the potential ${\mathbb V}(x)$ associated to ${\mathbb H}$ has to be determined from the 
very algebraic treatment.

In the realization just built $L^+$ and $L^-$ are differential ladder operators of order $m+1$.
Let us consider now the functions $\phi(x)$ which belong to the kernel of $L^-$,
\begin{equation}\label{0.4.4}
  L^-\phi=0\quad\Rightarrow \quad N({\mathbb H})\phi=L^+L^-\phi=\prod_{i=1}^{m+1}({\mathbb H} - \varepsilon_i)\phi=0.
\end{equation}
Since this kernel is invariant under ${\mathbb H}$, we can choose as the linearly independent functions $\phi$ generating 
this subspace the solutions of the stationary Schr\"odinger equation for ${\mathbb H}$ associated to $\varepsilon_i$:
\begin{equation}\label{0.4.6}
  {\mathbb H}\phi_{\varepsilon_i}=\varepsilon_i\phi_{\varepsilon_i}.
\end{equation}
Departing from each of these extremal states $\phi_{\varepsilon_i}$ it can be constructed a ladder of eigenfunctions of 
${\mathbb H}$ associated to the eigenvalues $\varepsilon_i + n, \ n=0,1,2,\dots$ so that the system described by ${\mathbb H}$ will have at most $m+1$ ladders with eigenfunctions built up by the repeated action of $L^+$ onto such extremal states.

By taking $m=0,1$, and looking for the more general systems ruled by the corresponding polynomial Heisenberg algebras, 
we will arrive to the harmonic oscillator and effective `radial' oscillator potentials (which have ladder operators of 
first and second orders respectively). On the other hand, for $m=2$ (third order ladder operators which will be 
specifically denoted by $l^\pm$ anticipating the reduced operators obtained from the theorem in section \ref{oscilador}) 
it turns out that the corresponding potential becomes determined by a function which satisfies the Painlev\'e IV equation 
\cite{acin00,cfnn04} (see also \cite{in04}). In order to see this explicitly, let us assume that
\begin{equation}
 {\mathbb H} = -\frac12 \frac{\rm d^2}{\rm d x^2} + {\mathbb V}(x), 
\end{equation}
and $\emph{l}^+=I_1^+I_2^+$, $\emph{l}^-=I_2^-I_1^-$, where
\begin{equation}\label{0.4.8}
I_1^+=\frac{1}{\sqrt{2}}\left[-\frac{\rm d}{\rm d x}+f(x)\right], \qquad I_2^+=\frac{1}{2}\left[\frac{\rm d^2}{\rm d x^2}+g(x)\frac{\rm d}{\rm d x}+h(x)\right].
\end{equation}
The previous factorized expressions for $l^\pm$ are useful since it is employed an auxiliar Hamiltonian 
$H_{\scriptsize \mbox{a}}=-\frac{1}{2}\frac{\rm d^2}{\rm d x^2}+ V_{\scriptsize \mbox{a}}(x)$ which is intertwined with ${\mathbb H}$ as follows 
\begin{equation}\label{0.5.9}
 ({\mathbb H} - 1)I_1^+=I_1^+H_{\scriptsize \mbox{a}}, \qquad H_{\scriptsize \mbox{a}} I_2^+=I_2^+{\mathbb H}. 
\end{equation}
By using then the formulae obtained for 1-SUSY and 2-SUSY and after several calculations we arrive to the following final
results:
\begin{eqnarray}\label{0.5.a1}
f & = & x + g, \\
\label{0.5.a3}
h & = & \frac{g'}{2} - \frac{g^2}{2} - 2xg - x^2 + \varepsilon_2 + \varepsilon_3 - 2 \varepsilon_1 - 1, \\
\label{0.5.a4}
\mathbb{V} & = & \frac{x^2}{2} - \frac{g'}{2} + \frac{g^2}{2} + xg + \varepsilon_1 - \frac{1}{2},
\end{eqnarray}
where $g(x)$ satisfies the Painlev\'e IV (PIV) equation,
\begin{equation}\label{0.5.10}
 g''=\frac{(g')^2}{2g}+\frac{3}{2}g^3+4xg^2+2(x^2-a)g+\frac{b}{g} ,
\end{equation}
with parameters $a=\varepsilon_2+\varepsilon_3-2\varepsilon_1-1$, $b=-2(\varepsilon_2 - \varepsilon_3)^2$.

Let us recall that $\varepsilon_i, \ i=1,2,3$ are the three roots involved in (\ref{0.4.3}) for $m=2$, which 
at the same time coincide with the energies for the three extremal $\phi_{\varepsilon_i}$ of ${\mathbb H}$, i.e.,
\begin{equation}\label{0.6.2}
 \emph{l}^-\phi_{\varepsilon_i} = 0 = N\phi_{\varepsilon_i} = \emph{l}^+\emph{l}^-\phi_{\varepsilon_i}, \qquad 
i = 1,2,3.
\end{equation}
Since $\emph{l}^-=I_2^-I_1^-$, where $I_2^-=(I_2^+)^\dagger$ and $I_1^-=(I_1^+)^\dagger$ arise from (\ref{0.4.8}), one of the extremal states, 
denoted $\phi_{\varepsilon_1}$, can be easily obtained  
\begin{equation}\label{0.6.3}
 I_1^-\phi_{\varepsilon_1} = \frac{1}{\sqrt{2}}\left[\frac{\rm d}{\rm d x}+f(x)\right]\phi_{\varepsilon_1}=0\quad \Rightarrow \quad   \phi_{\varepsilon_1} =  c \, \exp\left( - \frac{x^2}{2} - \int g \rm d x\right). 
\end{equation}
The other two extremal states are found in a more complicated way; however, their analytic expressions can
be obtained explicitly \cite{cfnn04}:
\begin{eqnarray}
&& \phi_{\varepsilon_2} \propto \left(\frac{g'}{2g} - \frac{g}{2} + \frac{\varepsilon_3 - \varepsilon_2}{g} - x\right)
\exp\left[\int\left(\frac{g'}{2g} + \frac{g}{2} + \frac{\varepsilon_3 - \varepsilon_2}{g}\right) \rm d x\right] , \\
&& \phi_{\varepsilon_3} \propto \left(\frac{g'}{2g} - \frac{g}{2} + \frac{\varepsilon_2 - \varepsilon_3}{g} - x\right)
\exp\left[\int\left(\frac{g'}{2g} + \frac{g}{2} + \frac{\varepsilon_2 - \varepsilon_3}{g}\right) \rm d x\right] .
\end{eqnarray}

Thus, given a solution $g$ of the Painlev\'e IV equation consistent with the parameter choice $a=\varepsilon_2 +
\varepsilon_3-2\varepsilon_1-1$, $b=-2(\varepsilon_2 - \varepsilon_3)^2$, $\varepsilon_{1,2,3}\in{\mathbb R}$, a system 
obeying a second-order polynomial Heisenberg algebra, characterized by the potential of (\ref{0.5.a4}), can be 
constructed. 

On the other hand, if we find a system ruled by second-order polynomial Heisenberg algebras, in particular 
its extremal states, then we can build solutions to the Painlev\'e IV equation as long as the extremal state is not identically null. 
In order to see this, let us rewrite the expression for the extremal state $\phi_{\varepsilon_1}$ of (\ref{0.6.3}) in the form:
\begin{equation}\label{0.6.4}
 g(x)=-x-[\ln\phi_{\varepsilon_1}]' ,
\end{equation}
i.e., a solution $g(x)$ to the PIV equation in terms of the extremal state $\phi_{\varepsilon_1}$ of ${\mathbb H}$ 
has been found. 

\subsection{Harmonic Oscillator}\label{oscilador}

Let us apply now the $k$-SUSY technique to the harmonic oscillator Hamiltonian
\begin{equation}\label{0.5.0}
  H=-\frac{1}{2}\frac{\rm d^2}{\rm d x^2}+\frac{x^2}{2}.
\end{equation}
If the transformation is of order $k\geq1$, it is possible to create $k$ new levels below the ground state energy $E_0 = 1/2$ of the
oscillator (let us suppose that this happens), at the positions defined by the {\it factorization energies} $\epsilon_j$, $j=1,..,k$ 
involved in (\ref{0.2.2}) \cite{fh99,ff05}. The eigenfunctions $\phi_n(x)$ of the new Hamiltonian $\tilde{H}$, associated to the eigenvalues 
$E_n = n + 1/2$ of the initial Hamiltonian $H$, are given by a generalization of (\ref{0.2.8b}) and (\ref{0.2.16}):
\begin{equation}
 \phi_n(x)=\frac{A^+\psi_n(x)}{\sqrt{(E_n-\epsilon_1)...(E_n-\epsilon_k)}}.
\end{equation}
Furthermore, the eigenfunctions $\phi_{\epsilon_j}$ associated to the new levels $\epsilon_j$ can be written as
\begin{equation}
 \phi_{\epsilon_j}\propto\frac{W(u_1,..,u_{j-1},u_{j+1},...,u_k)}{W(u_1,..,u_k)}, \quad j=1,..,k,
\end{equation}
where $W(u_1,..,u_k)$ is the Wronskian of the $k$ seed solutions $u_j$, $j=1,..,k$ used to implement the transformation,
which satisfy
\begin{equation}
 Hu_j=\epsilon_ju_j.
\end{equation}
Up to a constant factor, the general solution to this equation with $V(x)=\frac{x^2}{2}$ and $\epsilon$ arbitrary is given by 
\begin{equation}
 u(x)=\rm e^{-x^2/2} \left[\,_1 \mbox{F}_1\left(\frac{1-2\epsilon}{4};\frac{1}{2};x^2\right) + 2\nu \frac{\Gamma(\frac{3-2\epsilon}{4})}{\Gamma(\frac{1-2\epsilon}{4})} \ x\,_1 \mbox{F}_1\left(\frac{3-2\epsilon}{4};\frac{3}{2};x^2\right)\right]. 
\end{equation}
Thus, each $u_j$ takes this form with $\epsilon$ substituted by $\epsilon_j$ and $\nu$ by $\nu_j$. 
For this transformation not to be singular $W(u_1,..,u_k)$ must not have zeros in the real axis. 
For simplicity let us assume from now on that $\epsilon_k<\epsilon_{k-1}<...<\epsilon_1<E_0=1/2$. 
With this ordering $W(u_1,...,u_k)$ will not have zeros if $|\nu_j|<1$ for $j$ odd and $|\nu_j|>1$ for $j$ even, and thus the new potential
\begin{equation}
 \tilde{V}(x)=\frac{x^2}{2}-\left[\ln W(u_1,...,u_k) \right]''
\end{equation}
will not have singularities.

It is important to notice that the Hamiltonian $\tilde{H}$ has well defined ladder operators. In fact, the harmonic oscillator
has first order ladder operators $a^+$ and $a^-$. Let us define now two ladder operators for the system described by 
$\tilde{H}$ \cite{mi84,fh99}:
\begin{equation}\label{0.5.1}
 L^+=A^+a^+A, \qquad L^-=A^+a^-A.
\end{equation}
While $A^+$ and $A$ are differential operators of order $k$, $L^+$ and $L^-$ are of order $2k+1$. 
Due to the intertwining relations (\ref{0.1.2}) and the defining commutation relations of the ladder operators $a^+$ and $a^-$, 
$[H,a^\pm]=\pm a^\pm$, it turns out that $[\tilde{H},L^\pm ]=\pm L^\pm$, i.e., $L^+$ and $L^-$ are $(2k+1)$-th order ladder operators 
for $\tilde{H}$.

If $k=1$ the ladder operators $L^\pm$ are of third order and $\{\tilde{H},L^+,L^-\}$ directly generate a second order polynomial 
Heisenberg algebra. On the other hand, since $A^+$ and $A$ are of second order if $k=2$, then $L^\pm$
will be of fifth order in such a case. It is important to know under which circumstances $L^\pm$ can be `reduced' to third order 
ladder operators. The answer is contained in the following theorem \cite{bf11a}: if the seed solutions $u_1(x)$ and $u_2(x)$ are 
such that $u_2=a^-u_1$ and $\epsilon_2=\epsilon_1-1$, then $L^\pm$ can be factorized as 
\begin{equation}\label{0.5.11} 
 L^+ =\left(\tilde{H}-\epsilon_1\right)\emph{l}^+,\qquad L^- =\emph{l}^- \left(\tilde{H}-\epsilon_1\right),
\end{equation}
where $\emph{l}^+$ and $\emph{l}^-$ are third order differential ladder operators of $\tilde{H}$, such that  
$[\tilde{H},\emph{l}^\pm]=\pm \emph{l}^\pm$, which also satisfy
\begin{equation}
 \emph{l}^+\emph{l}^-=(\tilde{H}-\epsilon_2)(\tilde{H}-\epsilon_1-1)(\tilde{H}-1/2). 
\end{equation}

Once the Hamiltonian having third-order differential ladder operators is identified, it is straightforward to generate 
then the solutions to the Painlev\'e IV equation through its extremal states (see (\ref{0.6.4})). Using this technique, 
plenty of non-singular explicit solutions to the PIV equation, either real or complex, have been derived. It would 
be interesting to explore systematically the corresponding singular solutions. In this paper we will start this study 
by allowing the existence of one fixed singularity at the origin. Since our treatment is based on the harmonic oscillator 
with an infinite potential barrier at $x=0$, it is natural to start first by studying the associated problem of eigenvalues 
and then its corresponding SUSY partners.

\begin{figure}[h]  
\begin{center} 
  \includegraphics[width=5in]{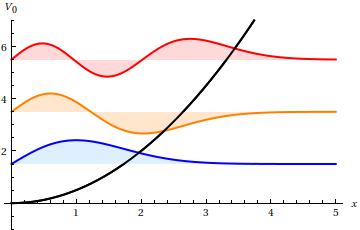} 
  \caption{\footnotesize The potential $V_{0}$ and its first three eigenfunctions. }
\label{gV0}
\end{center}
\end{figure}

\section{Harmonic oscillator with an infinite potential barrier at the origin}

We are interested in studying the Hamiltonian $H_0=-\frac{1}{2}\frac{\rm d^2}{\rm d x^2}+V_0(x)$ with
\[
  V_{0}(x) = \left\{
  \begin{array}{l l}
    \frac{x^{2}}{2} & \quad \text{if $x>0$  }\\
    \infty & \quad \text{if $x \leq 0.$ }\\
  \end{array} \right.
\]
The eigenvalues of $H_0$ take the form $E_n=2n+\frac{3}{2}$ with corresponding eigenfunctions
\begin{equation}\label{1.0i} 
 \psi_n(x)=C_n\,x\,\rm e^{-x^2/2}\,_1\mbox{F}_1\bigg(-n;\frac{3}{2};x^2\bigg),
\end{equation}
with $n\in\mathbb{N}$ and $C_n=2\frac{(2n+1)!}{\pi^{1/4}n!}\sqrt{\frac{2^{-2n}}{(2n+1)!}}$ being normalization constants \cite{fl71}.
These eigenfunctions $\psi_n(x)$ correspond to the odd eigenfunctions of the standard harmonic oscillator normalized in the domain 
$(0,\infty)$, which are the ones that satisfy the boundary conditions at $x=0$ and in the limit $x\rightarrow\infty$. 
A plot of the potential corresponding to the harmonic oscillator with an infinite potential barrier along with the first three 
eigenfunctions can be found in figure \ref{gV0}.

It will be required further ahead the even eigenfunctions of the standard harmonic oscillator but now normalized in the domain 
$(0,\infty)$,
\begin{equation}\label{1.0p}
 \chi_n(x)=B_n \, \rm e^{-x^2/2}\,_1\mbox{F}_1\bigg(-n;\frac{1}{2};x^2\bigg),
\end{equation}
which are associated to $\mathcal{E}_{n}=2n+ \frac12$, where $n\in\mathbb{N}$ and 
$B_n=\frac{(2n)!}{\pi^{1/4}n!}\sqrt{\frac{2^{1-2n}}{(2n)!}}$ are their normalization constants \cite{fl71}. 
Although they satisfy $H_0\chi_n=\mathcal{E}_n\chi_n$, they do not obey the boundary condition at $x=0$, and thus they 
are not eigenfunctions of $H_0$. We will say then that the $\chi_n$ are non-physical eigenfunctions (NPE)
of $H_0$ associated to $\mathcal{E}_{n}$.

\begin{figure}[h]  
\begin{center} 
  \includegraphics[width=5in]{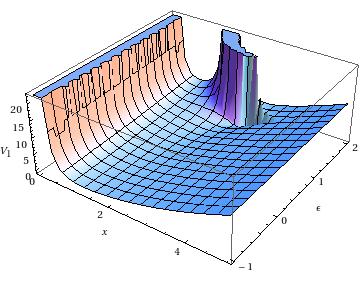}\\ 
  \caption{\footnotesize The potential $V_1(x)$ as a function of $x$ and the factorization energy $\epsilon$ for odd transformation functions. }
\label{gV1Ie}
\end{center}
\end{figure}

\subsection{1-SUSY}\label{1susy} 
Let us suppose now that $H_0$ is intertwined with another Hamiltonian $H_1=-\frac{1}{2}\frac{\rm d^2}{\rm d x^2}+V_1$ 
as in (\ref{0.1.2}), where $H_0$ is identified with the initial Hamiltonian $H$ and $H_1$ with the final 
one $\tilde H$ and the interwining operators $A^+, \ A$ are given by (\ref{0.1.3}). Consequently, the 
pair of Hamiltonians $H_0$ and $H_1$ can be factorized in the following way (see (\ref{0.2.2})):
\begin{equation}\label{1.H0} 
  H_{0}=AA^++\epsilon, \quad  H_{1}=A^+A+\epsilon,
\end{equation}
where the factorization energy $\epsilon$ is supposed to be real. In addition, the transformation function $u(x)$ must satisfy
the stationary Schr\"odinger equation:
\begin{equation}\label{1.6} 
  -\frac{1}{2}u''+V_0 u=\epsilon u.
\end{equation}
For $x>0$, this equation has a general solution given by 
\begin{equation}\label{1.6u}
  u(x)=\rm e^{-x^2/2} \left[ b_1\,_1 \mbox{F}_1\left(\frac{1-2\epsilon}{4};\frac{1}{2};x^2\right)+b_2\,x\,_1 \mbox{F}_1\left(\frac{3-2\epsilon}{4};\frac{3}{2};x^2\right)\right],
\end{equation}
$b_1,b_2$ being real constants \cite{jr98}. Since the bound states of $H_0$ vanish at $x=0$ and for $x\rightarrow\infty$, 
then the same boundary conditions will be required for the eigenfunctions of the new Hamiltonian $H_1$. This implies
that the transformation function of (\ref{1.6u}) must have a well defined parity in order that the bound states 
of $H_1$ vanish at $x=0$; hence, two different cases can be identified.

\begin{figure}[h]  
\begin{center} 
\includegraphics[width=5in]{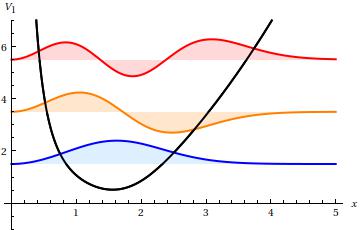}\\ 
\caption{\footnotesize The potential $V_{1}$ and its first three eigenfunctions obtained from an odd seed solution with factorization energy $\epsilon=\frac{1}{4}$. }
\label{gV1I}
\end{center}
\end{figure}

\subsection{Odd transformation function}

Let us choose in the first place an odd transformation function by taking $b_1=0$ and $b_2=1$ in (\ref{1.6u}) so that
\begin{equation}\label{1.u}
 u(x)=x \, \rm e^{-x^2/2}\,_1\mbox{F}_1\bigg(\frac{3-2\epsilon}{4};\frac{3}{2};x^2\bigg) .
\end{equation}
The substitution of this expression in (\ref{0.1.8a}) leads us immediately to the new potential
\begin{equation}\label{1.8} 
 V_1=V_0+\frac{1}{x^2}+1-\bigg\{\ln\left[_1\mbox{F}_1\bigg(\frac{3-2\epsilon}{4};\frac{3}{2};x^2\bigg)\right]\bigg\}'', \qquad  x>0.
\end{equation}
This potential contains a term of the form $\frac{1}{x^2}$, which is singular at $x=0$ and by itself induces 
in a natural way the vanishing boundary condition for the eigenfunctions of $H_1$ at $x=0$. Transformations 
with $\epsilon>\frac{3}{2}$ are not allowed since they generate additional singularities for $x>0$, and 
thus they modify the domain $(0,\infty)$ of the initial potential. Figure \ref{gV1Ie} shows the dependence of
$V_1(x)$ on $x$ and on the parameter $\epsilon$ for an intertwining that uses an odd transformation function.

As was shown in (\ref{0.2.8b}), an eigenfunction $\psi_n(x)$ of $H_0$ associated to the eigenvalue $E_n$ 
typically transforms into an eigenfunction $\phi_n(x)$ of $H_1$ associated to $E_n$, i.e.,
\begin{equation}\label{1.9}
 \phi_n(x)\propto A^+\psi_n(x)\propto-\psi_n'+\frac{u'}{u}\psi_n\propto\frac{W\left[u,\psi_n\right]}{u}.
\end{equation}

\noindent In our case this becomes true (see however the next subsection), and when substituting the expressions for $u(x)$ 
and $\psi_n(x)$ we obtain explicitly the eigenfunctions $\phi_n(x)$ (which satisfy the equation $H_1\phi_n=E_n\phi_n$ and
the boundary conditions $\phi_n(0) = \phi_n(\infty) = 0$):
\begin{eqnarray}\label{1phi}
 &\phi_n(x)=-D_nx^2{\rm e}^{-x^2/2}\Bigg\{\frac{4n}{3}\,_1\mbox{F}_1\Big(1-n;\frac{5}{2};x^2\Big)\quad\qquad \nonumber \\
 &\qquad\qquad\qquad\quad+\Big(1-\frac{2}{3}\epsilon\Big)\Bigg[\frac{\,_1\mbox{F}_1(\frac{7-2\epsilon}{4};\frac{5}{2};x^2)}{\,_1\mbox{F}_1(\frac{3-2\epsilon}{4};\frac{3}{2};x^2)}\Bigg]\,_1\mbox{F}_1\Big(-n;\frac{3}{2};x^2\Big)\Bigg\},
\end{eqnarray}
with $D_n=\frac{C_n}{\sqrt{2(E_n-\epsilon)}}$ being normalization constants. The corresponding energies $E_n = 2n+\frac32,
\ n=0,1,2,\dots$ thus belong to the spectrum of $H_1$. Some eigenfunctions $\phi_n(x)$ along with their corresponding potential
have been drawn in figure \ref{gV1I}.

The even eigenfunctions of the standard harmonic oscillator $\chi_n(x)$, which are NPE of $H_0$, transform 
into NPE $\varphi_n(x)$ of $H_1$ which diverge at $x=0$, as can be seen from the following explicit 
expressions which were calculated by using the right hand side of (\ref{1.9}) with $\psi_n$ substituted by $\chi_n$:
\begin{eqnarray}\label{1nofis}
 &\varphi_n(x)\propto \frac{1}{x}{\rm e}^{-x^2/2}\Bigg\{4nx^2\,_1\mbox{F}_1\Big(1-n;\frac{3}{2};x^2\Big)\qquad\qquad \nonumber\\ 
 &\qquad\qquad\qquad\qquad\quad+\Bigg[(1-\frac{2}{3}\epsilon)x^2\frac{\,_1\mbox{F}_1(\frac{7-2\epsilon}{4};\frac{5}{2};x^2)}{\,_1\mbox{F}_1(\frac{3-2\epsilon}{4};\frac{3}{2};x^2)}+1\Bigg]\,_1\mbox{F}_1\Big(-n;\frac{1}{2};x^2\Big)\Bigg\}.
\end{eqnarray} 
Thus, the energies $\mathcal{E}_n = 2n+\frac12$, $n=0,1,2,...$ do not belong to the spectrum of $H_1$. 
In addition, since $\phi_\epsilon(x)\propto 1/u(x)$ diverges also for $x=0$, then $\epsilon$ belongs neither to the spectrum of $H_1$.

The limit case $\epsilon\rightarrow\frac{3}{2}$ is worth of attention, 
since for this factorization energy the ground state energy level of $H_0$ is erased from the spectrum of the new hamiltonian $H_1$. 
Although the new spectrum is equivalent to the old one through a finite displacement in the energy, the form of the new potential, 
however, is drastically different from the initial one due to the singular term $1/x^2$ (see (\ref{1.8})).

\begin{figure}[h]  
\begin{center} 
  \includegraphics[width=5in]{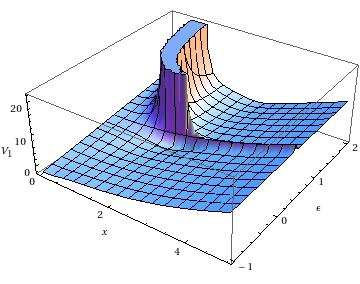}\\ 
  \caption{\footnotesize The potential $V_1(x)$ as a function of $x$ and the factorization energy $\epsilon$ for even transformation functions. }
\label{gV1Pe}
\end{center}
\end{figure}

\subsection{Even transformation function}

Let us choose now the even solution of (\ref{1.6}) as transformation function, 
\begin{equation}\label{1up}
  u(x)=\rm e^{-x^2/2} \,_1 \mbox{F}_1\left(\frac{1-2\epsilon}{4};\frac{1}{2};x^2\right).
\end{equation}
Using once again (\ref{0.1.8a}), the potential $V_1(x)$ turns out to be
\begin{equation}\label{1.12}
 V_1(x)=V_0(x)+1-\Bigg\{\ln\left[_1\mbox{F}_1\left(\frac{1-2\epsilon}{4};\frac{1}{2};x^2\right)\right]\Bigg\}''.
\end{equation}
This potential has also a singularity at $x=0$, since $V_0(x)$ includes the infinite potential barrier.
Transformations with $\epsilon>\frac{1}{2}$ are not allowed, due to they generate additional singularities 
for $x>0$ and, thus, they modify the domain of definition of the initial potential. Figure \ref{gV1Pe} shows 
the dependence of $V_1(x)$ on $x$ and on the parameter $\epsilon$ for an intertwining that uses an even transformation function.

\begin{figure}[h]  
\begin{center} 
  \includegraphics[width=5in]{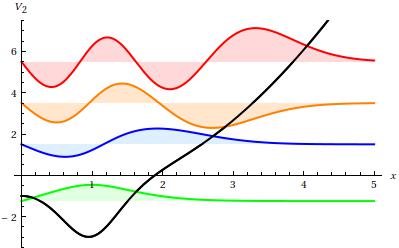}\\ 
  \caption{\footnotesize The potential $V_{1}$ and its first three eigenfunctions obtained from an even seed solution with factorization energy $\epsilon=\frac{1}{4}$. }
\label{gV1P}
\end{center}
\end{figure}

Let us calculate now the eigenfunctions of $H_1$ from the corresponding ones of $H_0$ using (\ref{1.9}). 
For the odd eigenfunctions $\psi_n(x)$ we obtain
\begin{eqnarray}\label{1.13}
 &\varphi_n(x)\propto {\rm e}^{\frac{-x^2}{2}}\Bigg\{4nx^2\,_1\mbox{F}_1\Big(1-n;\frac{5}{2};x^2\Big)\qquad\qquad\quad \nonumber\\ 
 &\qquad\qquad\qquad\qquad\qquad+3\Bigg[\Big(1-2\epsilon\Big)\,x^2\,\frac{\,_1\mbox{F}_1(\frac{5-2\epsilon}{4};\frac{3}{2};x^2)}{\,_1\mbox{F}_1(\frac{1-2\epsilon}{4};\frac{1}{2};x^2)}-1\Bigg]\,_1\mbox{F}_1\Big(-n;\frac{3}{2};x^2\Big)\Bigg\}, 
\end{eqnarray}
where the symbol $\varphi$ is used since they are NPE of $H_1$ associated to $E_n$. On the other hand, 
the even NPE $\chi_n(x)$ of $H_0$ are now mapped into the correct eigenfunctions of $H_1$:
\begin{eqnarray}\label{1.14}
 &\phi_n(x)=D_n x \, {\rm e}^{\frac{-x^2}{2}}\Bigg\{4n\,_1\mbox{F}_1\Big(1-n;\frac{3}{2};x^2\Big)\qquad\qquad \nonumber \\
 &\qquad\qquad\quad+\Big(1-2\epsilon\Big)\frac{\,_1\mbox{F}_1(\frac{5-2\epsilon}{4};\frac{3}{2};x^2)}{\,_1\mbox{F}_1(\frac{1-2\epsilon}{4};\frac{1}{2};x^2)}\,_1\mbox{F}_1\Big(-n;\frac{1}{2};x^2\Big)\Bigg\}, 
\end{eqnarray}
with $D_n=\frac{B_n}{\sqrt{2(\mathcal{E}_n-\epsilon)}}$ being normalization constants. Some of them are plotted in figure \ref{gV1P} along with the corresponding potential. Note that $\phi_n(x)$
satisfy the boundary conditions, while $\varphi_n(x)$ do not. As in the previous case, the function 
$\phi_\epsilon(x)\propto 1/u(x)$ does not obey the boundary condition at $x=0$ and thus the associated factorization energy 
$\epsilon$ does not belong to the spectrum of $H_1$. In conclusion, the spectrum of $H_1$ is composed by the levels 
${\cal E}_n = 2n + \frac12, \ n=0,1,\dots$

Once again, there is a notorious limit $\epsilon\rightarrow\frac{1}{2}$, since in this case the otherwise ground state energy level ${\cal E}_0=\frac{1}{2}$ is erased from the spectrum of the new hamiltonian $H_1$. Notice that in this case the new potential and its spectrum become the same as the initial ones (up to a finite displacement in the energy).

\subsection{2-SUSY}\label{2susy}

Let us suppose that $H_0$ is intertwined with a different Hamiltonian $H_2$ as in (\ref{0.1.2}), 
$H_2$ being identified now with $\tilde H$ and the intertwining operators $A^+, \ A$ with the second order ones of 
(\ref{0.1.9}). According to subsection \ref{esegundo}, the transformation functions $u_1(x)$ and $u_2(x)$ 
must satisfy the stationary Schr\"odinger equation, whose general solution for $x>0$ is the one 
of (\ref{1.6u}). From (\ref{0.1.28}) we can see that the new potential can be written as
\begin{equation}
  V_2=V_0-\eta'=V_0- \left[\ln W\left(u_1,u_2\right)\right]''.
\end{equation}
In addition, the eigenfunctions $\psi_n(x)$ of $H_0$ typically transform into eigenfunctions $\phi_n(x)$ of 
$H_2$ through the action of the intertwining operator $A^+$ as follows:
\begin{eqnarray}
  &\phi_n(x)=\frac{A^+\psi_n(x)}{\sqrt{(E_n-\epsilon_1)(E_n-\epsilon_2)}}\qquad\qquad\qquad\qquad\qquad\qquad\qquad\quad\quad \nonumber \\
 &\qquad\, \propto - \left[\ln W\left( u_1,u_2  \right)\right]'\psi_n'(x)\qquad\qquad\qquad\qquad\qquad\qquad \nonumber \\
 &\qquad\qquad\qquad\qquad   + \left( \frac{\left[\ln W\left( u_1,u_2  \right)\right]''}{2}+\frac{\big\{\left[\ln W\left( u_1,u_2  \right)\right]'\big\}^2}{2}-2E_n+\epsilon_1+\epsilon_2 \right)\psi_n(x).
\end{eqnarray}

As it was seen previously, for a given $\epsilon$ there are two solutions with opposite parity: the odd solution 
$u(x) = x \, \rm e^{-\frac{x^2}{2}}\,_1 \mbox{F}_1(\frac{3-2\epsilon}{4};\frac{3}{2};x^2)$ and the even one 
$u(x)=\rm e^{-\frac{x^2}{2}}\,_1 \mbox{F}_1(\frac{1-2\epsilon}{4};\frac{1}{2};x^2)$. We are going to choose $u_1(x)$ and 
$u_2(x)$ as parity definite solutions for the ordering $\epsilon_1 > \epsilon_2$. 
Note that there exist four different parity combinations leading to four different kinds of second-order 
transformations which are explored below. 

\begin{figure}[h]  
\begin{center} 
\includegraphics[width=5in]{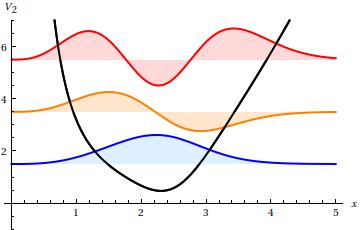}\\ 
\caption{\footnotesize The potential $V_{2}$ and its first three eigenfunctions obtained from two odd seed solutions 
 with factorization energies $\epsilon_1=\frac{3}{8}$ and $\epsilon_2=\frac{1}{8}$. }
\label{gV2II}
\end{center}
\end{figure}

\subsection{Odd-odd transformation functions}

Let us choose $u_1=x\,\rm e^{-\frac{x^2}{2}}\,_1 \mbox{F}_1(\frac{3-2\epsilon_1}{4};\frac{3}{2};x^2)$ and
$u_2=x\,\rm e^{-\frac{x^2}{2}}\,_1 \mbox{F}_1(\frac{3-2\epsilon_2}{4};\frac{3}{2};x^2)$.
Since $u_1(x)=F(x)G(x)$ and $u_2(x)=F(x)H(x)$ with $F(x)=x\,\rm e^{-x^2/2}$, 
$G(x)=\,_1\mbox{F}_1\left(\frac{3-2\epsilon_1}{4};\frac{3}{2};x^2 \right)$,
$H(x)=\,_1\mbox{F}_1\left(\frac{3-2\epsilon_2}{4};\frac{3}{2};x^2 \right)$, then
the wronskian $W(u_1,u_2)$ can be expressed as
\begin{equation}\label{wnosingular}
W(u_1,u_2)=W(FG,FH)=F^2W(G,H)=x^3\rm e^{-x^2}w(x),
\end{equation}
where $w(x)\equiv\frac{W(G,H)}{x}$ turns out to be a continuous function without zeros in $x\geq0$.
In this way we have separated the singularity at $x=0$ induced by the transformation on the new potential 
(e.g. \cite{cf08}), i.e., 
\begin{equation}
 V_2(x)=V_0-\{\ln\left[W(u_1,u_2)\right]\}''=\frac{x^2}{2}+\frac{3}{x^2}+2-\left[\ln w(x)\right]''\quad\text{for}\quad x\geq0.
\end{equation}

The eigenfunctions $\phi_n(x)$ of $H_2$ can be found from those of $H_0$ in the standard way, 
$\phi_n(x)=\frac{A^+\psi_n(x)}{\sqrt{(E_n-\epsilon_1)(E_n-\epsilon_2)}}$, and they satisfy the appropriate 
boundary conditions so that the eigenvalues $E_n$ in general belong to the spectrum of $H_2$.
On the other hand, the even NPE $\chi_n(x)$ of $H_0$ given by (\ref{1.0p}), 
transform into $\varphi_n(x)=\frac{A^+\chi_n}{\sqrt{(\mathcal{E}_n-\epsilon_1)(\mathcal{E}_n-\epsilon_2)}}$. 
These last do not satisfy the boundary conditions at $x=0$ either, thus the corresponding energies $\mathcal{E}_n$ 
are not in the spectrum of $H_2$. In addition, the expressions for $u_1, \ u_2$ and the $W(u_1,u_2)$ of 
(\ref{wnosingular})  substituted in (\ref{twomissing}) show that $\phi_{\epsilon_1}$ and $\phi_{\epsilon_2}$
diverge at $x=0$ and, hence, neither $\epsilon_1$ nor $\epsilon_2$ belong to the spectrum of $H_2$.

It is worth noticing that this choice of $u_1(x)$ and $u_2(x)$ produces a non-singular transformation for $x>0$ as long as the factorization energies satisfy $\epsilon_2 < \epsilon_1 \le\frac{3}{2}=E_0$ or $E_j=\frac{3+4j}{2}\le\epsilon_2 < \epsilon_1 \le \frac{3+4(j+1)}{2}=E_{j+1}$. As in the previous section, singular transformations are discarded since they modify the domain of definition of the initial potential and thus its corresponding spectral problem.

There are several limit cases through which we can delete either one or two levels of $H_0$ for arriving to $H_2$.
For instance, the initial ground state energy $E_0$ can be deleted by making $\epsilon_1 = E_0, \ \epsilon_2 < E_0$ 
since now the solution of the stationary Schr\"odinger equation for $H_2$ associated to $\epsilon_1 = E_0$ does not satisfy the boundary conditions 
and thus $E_0 \not\in {\rm Sp}(H_2)$. On the other hand, in the domain $E_j\le\epsilon_2 < \epsilon_1 \le E_{j+1}$ we 
can delete either $E_j$ or $E_{j+1}$, by taking $\epsilon_2 = E_j$ with $E_j<\epsilon_1 < E_{j+1}$ in the first case or 
$\epsilon_1 =E_{j+1}$ and $E_j<\epsilon_2 < E_{j+1}$ in the second. Moreover, the two consecutive levels $E_{j}, E_{j+1}$ can be deleted by choosing $\epsilon_2 = E_j$ and $\epsilon_1 = E_{j+1}$.

In figure \ref{gV2II} we can see an example of the new potential $V_2$ and several of its eigenfunctions $\phi_n(x)$ for 
$\epsilon_2 < \epsilon_1 < 3/2$.

\begin{figure}[h]  
\begin{center} 
\includegraphics[width=5in]{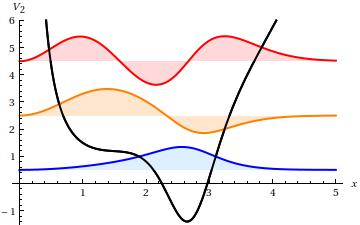}\\ 
\caption{\footnotesize The potential $V_{2}$ and its first three eigenfunctions obtained from two even seed solutions 
with factorization energies $\epsilon_1=\frac{3}{8}$ and $\epsilon_2=\frac{1}{8}$. }
\label{gV2PP}
\end{center}
\end{figure}

\subsection{Even-even transformation functions}

Let us take now $u_1=\rm e^{-\frac{x^2}{2}}\,_1 \mbox{F}_1(\frac{1-2\epsilon_1}{4};\frac{1}{2};x^2)$ and
$u_2=\rm e^{-\frac{x^2}{2}}\,_1 \mbox{F}_1(\frac{1-2\epsilon_2}{4};\frac{1}{2};x^2)$. We obtain 
that $W(u_1,u_2)=x\,\rm e^{-x^2}w(x)$, where $w(x)\equiv\frac{W(F,G)}{x}$ is a continous 
function without zeros for $x\geq 0$ (with $F=\,_1\mbox{F}_1(\frac{1-2\epsilon_1}{4};\frac{1}{2};x^2)$
and $G=\,_1\mbox{F}_1(\frac{1-2\epsilon_2}{4};\frac{1}{2};x^2)$). Hence
\begin{equation}
 V_2(x)=\frac{x^2}{2}+\frac{1}{x^2}+2-\left[\ln w(x)\right]''\quad\text{for}\quad x\geq0. 
\end{equation}

Note that the eigenfunctions $\psi_n$ of $H_0$ are mapped here into NPE
$\varphi_n(x)=\frac{A^+\psi_n(x)}{\sqrt{(E_n-\epsilon_1)(E_n-\epsilon_2)}}$ of $H_2$ that do not satisfy 
the boundary conditions and then the energies $E_n$ are not in the spectrum of $H_2$. Meanwhile, the 
even NPE $\chi_n(x)$ of $H_0$, that do not satisfy the boundary conditions, 
transform into the correct eigenfunctions $\phi_n(x)=\frac{A^+\chi_n}{\sqrt{(\mathcal{E}_n-\epsilon_1)(
\mathcal{E}_n-\epsilon_2)}}$ of $H_2$, which do satisfy the boundary conditions and thus, the corresponding 
eigenvalues $\mathcal{E}_n$ belong to the spectrum of $H_2$. As in the previous case, the NPE
$\phi_{\epsilon_{1,2}}$ of $H_2$ associated to $\epsilon_{1,2}$ diverge at $x=0$ and thus $\epsilon_{1,2}
\not\in{\rm Sp}(H_2)$. 

For this parity combination of $u_1$ and $u_2$ the transformation is non-singular for $x>0$ as long as the 
factorization energies satisfy $\epsilon_2<\epsilon_1\le\frac{1}{2} = {\cal E}_0$ or ${\cal E}_j=\frac{1+4j}{2}\le
\epsilon_2<\epsilon_1\le\frac{1+4(j+1)}{2} = {\cal E}_{j+1}$. Similarly to the previous section, singular 
transformations with singularities at $x>0$ are not allowed due to they change the initial spectral problem.

The limit cases for which one or two neighbour levels ${\cal E}_j$ disappear from ${\rm Sp}(H_2)$ work
similarly as in the previous case. Thus, by taking $\epsilon_1 = {\cal E}_0$, $\epsilon_2 < {\cal E}_0$
it turns out that ${\cal E}_0\not\in{\rm Sp}(H_2)$. On the other hand, if we make either $\epsilon_2 = 
{\cal E}_j$ with ${\cal E}_j<\epsilon_1<{\cal E}_{j+1}$ or $\epsilon_1 = {\cal E}_{j+1}$ with 
${\cal E}_j<\epsilon_2<{\cal E}_{j+1}$, it turns out that either ${\cal E}_j\not\in{\rm Sp}(H_2)$
or ${\cal E}_{j+1}\not\in{\rm Sp}(H_2)$ respectively. In addition, if $\epsilon_2 = {\cal E}_j$ and 
$\epsilon_1 = {\cal E}_{j+1}$  then both ${\cal E}_j, {\cal E}_{j+1}\not\in{\rm Sp}(H_2)$.

In figure \ref{gV2PP} one can find some examples of the eigenfunctions $\phi_n(x)$ 
along with the corresponding potential $V_2$ for $\epsilon_2<\epsilon_1<\frac{1}{2}$.

\begin{figure}[h]  
\begin{center} 
\includegraphics[width=5in]{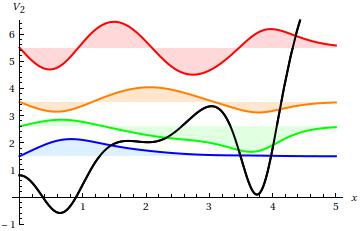}\\ 
\caption{\footnotesize The potential $V_{2}$ and its first four eigenfunctions obtained from odd and even seed solutions 
with factorization energies $\epsilon_1=\frac{6}{2}$ and $\epsilon_2=\frac{5}{2}+0.1$. }
\label{gV2IP}
\end{center}
\end{figure}

\subsection{Odd-even transformation functions}

Let $u_1=x\,\rm e^{-\frac{x^2}{2}}\,_1 \mbox{F}_1(\frac{3-2\epsilon_1}{4};\frac{3}{2};x^2)$ and
$u_2=\rm e^{-\frac{x^2}{2}}\ _1 \mbox{F}_1(\frac{1-2\epsilon_2}{4};\frac{1}{2};x^2)$ with
$\epsilon_2<\epsilon_1$. Since again $u_1(x)=F(x)G(x)$, $u_2(x)=F(x)H(x)$ with $F(x)=\rm e^{-x^2/2}$,
$G(x)=x\,_1\mbox{F}_1\left(\frac{3-2\epsilon_1}{4};\frac{3}{2};x^2\right)$ \ and \
$H(x)=\,_1\mbox{F}_1\left(\frac{1-2\epsilon_2}{4};\frac{1}{2};x^2\right)$, \ 
it turns out that now the Wronskian becomes $W(u_1,u_2)=\rm e^{-x^2}W(G,H)$, where $W(G,H)$ is a continuous function
without zeros for $x\geq 0$. Thus, the new potential can be written as:
\begin{equation}\label{2.17}
 V_2(x)=\frac{x^2}{2}+2-\left[\ln W(G,H)\right]''\quad\text{for}\quad x\geq0.
\end{equation}

Let us note that the eigenfunctions of $H_2$ are found here by acting the intertwining operator $A^+$
onto the those $\psi_n$ of $H_0$, 
\mbox{$\phi_n(x)=\frac{A^+\psi_n(x)}{\sqrt{(E_n-\epsilon_1)(E_n-\epsilon_2)}}$},
since they  satisfy the boundary conditions so that their corresponding eigenvalues $E_n$ belong to the spectrum of $H_2$.
Meanwhile, the even NPE $\chi_n$ of $H_0$, which do not satisfy the boundary conditions, transform into
NPE $\varphi_n(x)=\frac{A^+\chi_n(x)}{\sqrt{(\mathcal{E}_n-\epsilon_1)(\mathcal{E}_n-\epsilon_2)}}$
of $H_2$ that do not satisfy the boundary conditions and, consequently, the energies $\mathcal{E}_n$ 
do not belong to the spectrum of $H_2$. 

For this choice of $u_1(x)$ and $u_2(x)$ the transformation is found to be non-singular for $x>0$ as long as the factorization energies satisfy ${\cal E}_j = \frac{1+4j}{2}\le\epsilon_2 <\epsilon_1\le\frac{3+4j}{2} = E_j$. As in the previous section, singular transformations with singularities at $x>0$ are once again discarded.

Now we need to know if either $\phi_{\epsilon_1}$, $\phi_{\epsilon_2}$ or both in (\ref{twomissing})
satisfy the boundary conditions to become also eigenfunctions of $H_2$. For 
${\cal E}_j < \epsilon_2 <\epsilon_1 < E_j$ it turns out that $\phi_{\epsilon_2}$ satisfies the boundary conditions
while $\phi_{\epsilon_1}$ does not. This implies that $\epsilon_2\in {\rm Sp}(H_2)$ and $\epsilon_1\not\in 
{\rm Sp}(H_2)$, i.e., through the second-order SUSY transformation it can be created a new level at the position 
$\epsilon_2$. This is a surprising result since by means of the first-order SUSY transformation we produced potentials
which were just isospectral to the initial one. In addition, for $\epsilon_1 = E_j$ with 
${\cal E}_j < \epsilon_2 < E_j$ the same result is obtained, but now it implies that $\epsilon_1 = E_j \not\in 
{\rm Sp}(H_2)$ and $\epsilon_2\in {\rm Sp}(H_2)$. Thus, by employing the second-order SUSY transformation we have deleted 
the level $E_j$ and at the same time we have created a new one at $\epsilon_2$, so we have effectively `moved down'
$E_j$ to its new position $\epsilon_2$. For $\epsilon_2 = {\cal E}_j$ and ${\cal E}_j < \epsilon_1 < E_j$
neither $\phi_{\epsilon_1}$ nor $\phi_{\epsilon_2}$ satisfy the boundary conditions so that $\epsilon_{1,2}
\not\in{\rm Sp}(H_2)$. Finally, for $\epsilon_1 = E_j$ and $\epsilon_2 = {\cal E}_j$ the same happens, i.e.,
we have deleted the level $E_j$ in order to produce $H_2$.

In figure \ref{gV2IP} one can find an example of the potential $V_2$ along with some of its eigenfunctions 
$\phi_n(x)$ for ${\cal E}_1 < \epsilon_2 <\epsilon_1 < E_1$.

\begin{figure}[h]  
\begin{center} 
\includegraphics[width=5in]{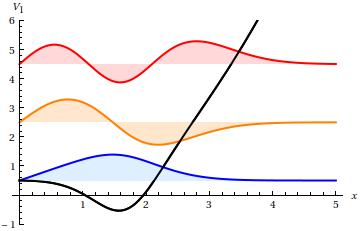} 
\caption{\footnotesize The potential $V_{2}$ and its first four eigenfunctions obtained from even and odd seed solutions 
with factorization energies $\epsilon_1=-\frac{5}{4}$ and $\epsilon_2=-\frac{7}{4}$.}
\label{gV2PI}
\end{center}
\end{figure}

\subsection{Even-odd transformation functions}

Finally let us take $u_1=\rm e^{-\frac{x^2}{2}}\,_1 \mbox{F}_1(\frac{1-2\epsilon_1}{4};\frac{1}{2};x^2)$ \ and \
$u_2=x\,\rm e^{-\frac{x^2}{2}}\,_1 \mbox{F}_1(\frac{3-2\epsilon_2}{4};\frac{3}{2};x^2)$ with \mbox{$\epsilon_2 <\epsilon_1$}.
A similar calculation as in the previous section leads to a $V_2(x)$ having the same form of (\ref{2.17}), 
where now $G(x)=\,_1 \mbox{F}_1(\frac{1-2\epsilon_1}{4};\frac{1}{2};x^2)$ and $H(x)=x\,_1 \mbox{F}_1(\frac{3-2\epsilon_2}{4};\frac{3}{2};x^2)$. 
Once again, the eigenfunctions $\phi_n$ of $H_2$ are obtained from those of $H_0$ through 
$\phi_n(x)=\frac{A^+\psi_n}{\sqrt{(E_n-\epsilon_1)(E_n-\epsilon_2)}}$, which satisfy the boundary conditions so that 
the eigenvalues $E_n$ belong to the spectrum of $H_2$. On the other hand, the even NPE $\chi_n(x)$ of $H_0$ 
which do not obey the boundary conditions of the original problem, transform into NPE
$\varphi_n(x)=\frac{A^+\chi_n}{\sqrt{(\mathcal{E}_n-\epsilon_1)(\mathcal{E}_n-\epsilon_2)}}$ of $H_2$ 
which do not satisfy neither the boundary conditions. Thus, their corresponding energies $\mathcal{E}_n$ are not contained in the 
spectrum of $H_2$.  

Note that for this choice of $u_1(x)$ and $u_2(x)$ the transformation is non-singular as long as the factorization energies
satisfy $\epsilon_2<\epsilon_1\le\frac{1}{2} = {\cal E}_0$ or $E_j=\frac{3+4j}{2}\le\epsilon_2<\epsilon_1\le\frac{5+4j}{2}={\cal E}_{j+1}$. 
As in the previous section, singular transformations are discarded since they modify the initial domain of the potential. 

By studying once again if the $\phi_{\epsilon_{1,2}}$ of (\ref{twomissing}) satisfy the boundary conditions 
we arrive now to the following results: for $\epsilon_2<\epsilon_1< {\cal E}_0$ or $E_j<\epsilon_2<\epsilon_1<{\cal E}_{j+1}$ 
it turns out that $\phi_{\epsilon_1}$ satisfies the boundary conditions while $\phi_{\epsilon_2}$ does not, 
meaning that $\epsilon_1\in {\rm Sp}(H_2)$ and $\epsilon_2\not\in{\rm Sp}(H_2)$, i.e., a new level has been created at
$\epsilon_1$. For $\epsilon_1 = {\cal E}_0$ and $\epsilon_2 < {\cal E}_0$ it is obtained that $\epsilon_{1,2}\not\in{\rm Sp}(H_2)$, namely,
there is no additional level in ${\rm Sp}(H_2)$. On the other hand, for $\epsilon_2=E_j$ and $E_j<\epsilon_1<{\cal E}_{j+1}$ 
once again $\epsilon_1\in {\rm Sp}(H_2)$ and $\epsilon_2=E_j\not\in{\rm Sp}(H_2)$, i.e., 
through the second-order SUSY transformation the level $E_j$ has been `moved up' to the position $\epsilon_1$. 
For $\epsilon_1={\cal E}_{j+1}$ and $E_j<\epsilon_2<{\cal E}_{j+1}$ neither 
$\phi_{\epsilon_1}$ nor $\phi_{\epsilon_2}$ satisfy the boundary conditions so that $\epsilon_{1,2}\not\in{\rm Sp}(H_2)$.
Finally, for $\epsilon_1={\cal E}_{j+1}$ and $\epsilon_2=E_j$ the same happens, which implies that the level $E_j$ has been deleted.

Figure \ref{gV2PI} shows a potential $V_2$ and some of its eigenfunctions $\phi_n(x)$ for 
$\epsilon_2<\epsilon_1<\frac{1}{2}$.

\section{Solutions to the Painlev\'e IV equation}

In section \ref{SQM} it was stated that it is possible to find solutions $g(x)$ to the Painlev\'e IV equation through
\begin{equation}
 g(x)=-x-[\ln\phi_{\varepsilon_1}]'
\end{equation}
where $\phi_{\varepsilon_1}$ is an extremal state for a system having third order ladder operators $\emph{l}^+$ and $\emph{l}^-$,
which satisfy (\ref{0.6.2}). Moreover, if we know the explicit form of the three extremal states 
(and their associated eigenvalues), we can identify  each one of these with $\phi_{\varepsilon_1}$ and thus three solutions to 
the PIV equation can be generated, associated to different parameters $a$, $b$. Since Hamiltonians generated from 
the harmonic oscillator with an infinite potential barrier at the origin through supersymmetric techniques can have third order 
ladder operators, hence solutions to the PIV equation can be straightforwardly obtained, as detailed ahead.
 
\subsection{1-SUSY}

Recall that for a first order supersymmetry transformation the analogue to the number operator is given by
\begin{equation}
 N \equiv L^+L^-=A^+a^+AA^+a^-A =\left(H_1-\epsilon\right)\left(H_1-1-\epsilon\right)\left(H_1-\frac{1}{2}\right).
\end{equation}
In addition, there are three extremal states $\phi_{1}$, $\phi_{2}$ and $\phi_{3}$ 
with eigenvalues $\varepsilon_1$, $\varepsilon_2$ and $\varepsilon_3$ respectively which satisfy
\begin{equation}\label{11}
 N\phi_{i}=L^+L^-\phi_{i}=0, \qquad i =1,2,3.
\end{equation}
Explicit expressions for such extremal states are well known, and we can label them firstly in the way:
\begin{equation}
 \phi_{1}\propto \frac{1}{u(x)}, \quad \phi_{2}\propto A^+a^+u(x), \quad \phi_{3}\propto A^+\chi_0,
\end{equation}
where $\{\varepsilon_1=\epsilon,\varepsilon_2=\epsilon+1,\varepsilon_3=\frac{1}{2}\}.$
Moreover, the cyclic permutations of the indices of $\{\varepsilon_1,\varepsilon_2,\varepsilon_3\}$
lead immediately to additional solutions of the PIV equation with parameters determined by
three different choices:
\{$\varepsilon_1=\epsilon$, $\varepsilon_2=\epsilon+1$, $\varepsilon_3=\frac{1}{2}$\}, 
\{$\varepsilon_1=\epsilon+1$, $\varepsilon_2=\frac{1}{2}$, $\varepsilon_3=\epsilon$\}, 
\{$\varepsilon_1=\frac{1}{2}$, $\varepsilon_2=\epsilon$, $\varepsilon_3=\epsilon+1$\}.

It is worth noticing that the solutions to PIV depend on our selection of the transformation function
$u(x)$, for which there are two different choices (for a fixed $\epsilon$).

\subsection{Odd transformation function}

For $u(x)=x\,\rm e^{-x^2/2}\,_1 \mbox{F}_1(\frac{3-2\epsilon}{4};\frac{3}{2};x^2)$ we obtain the following extremal states:
\begin{eqnarray}
\phi_{1}&\propto& \frac{\,\rm e^{x^2/2}}{x\,_1\mbox{F}_1(\frac{3-2\epsilon}{4};\frac{3}{2};x^2)}, \label{Iext2} \\
\phi_{2}&\propto& u[(\ln u)''-1], \label{Iext3} \\
\phi_{3}&\propto& \rm e^{-x^2/2}\Bigg[\Big(1-\frac{2}{3}\epsilon\Big)x\frac{\,_1\mbox{F}_1(\frac{7-2\epsilon}{4};\frac{5}{2};x^2)}{\,_1\mbox{F}_1(\frac{3-2\epsilon}{4};\frac{3}{2};x^2)}+\frac{1}{x}\Bigg]. \label{Iext1}
\end{eqnarray}
These expressions and the cyclic permutations of $\{\varepsilon_1,\varepsilon_2,\varepsilon_3\}$ lead to the following 
three solutions $g_{i}(x)=-x-[\ln\phi_{i}]'$ of the PIV equation
\begin{eqnarray}
g_{1} &=&  \frac{1}{x} - 2x + \left(1-\frac23\epsilon\right) x \, \frac{\,_1\mbox{F}_1(\frac{7-2\epsilon}{4};\frac{5}{2};x^2)}{\,_1\mbox{F}_1(\frac{3-2\epsilon}{4};\frac{3}{2};x^2)},\\
g_{2} &=& - g_1 - 2 x - 2 \left[\frac{x + (2\epsilon - x^2)(g_1+x) + (g_1+x)^3}{x^2-2\epsilon-1-(g_1+x)^2} \right],\\ 
g_{3} &=& - \frac{g_1'+2}{g_1 + 2x} = \frac{g_1^2 + 2 x g_1 + 2 \epsilon - 1}{g_1 + 2x} .
\end{eqnarray}

Note that $g_{1}$ solves the PIV equation with parameters $a_1=-\epsilon+\frac{1}{2}$ and $b_1=-2\left(\epsilon+\frac{1}{2}\right)^2$
while $g_{2}$ and $g_{3}$ do it for $a_2=-\epsilon-\frac{5}{2}$ and $b_2=-2\left(\epsilon-\frac{1}{2}\right)^2$ 
and $a_3=2\epsilon-1$ and $b_3=-2$ respectively.
Since they involve the confluent hypergeometric function, it is said that these belong to the 
confluent hypergeometric function hierarchy of solutions to PIV. In addition, for some particular values of the 
factorization energy $\epsilon$ they reduce to well known special functions, some examples of which 
are reported in the following table (here and in the following section $F(x)=\frac{1}{2}\sqrt{\pi}\,\rm e^{-x^2}erfi(x)$ will represent the Dawson function):
\\
\begin{center}
\begin{tabular}{c|ccc}
  & $\epsilon=-\frac{1}{2}$ & $\epsilon=\frac{1}{2}$ & $\epsilon=\frac{3}{2}$ \\ 
 \hline \\
 $g_{1}(x)$& $\frac{2\rm e^{-x^2}}{\sqrt{\pi}erf(x)}$ & $\frac{2\rm e^{-x^2}}{\sqrt{\pi}erfi(x)}-2x$ & $\frac{1}{x}-2x$ \\ \\
 $g_{2}(x)$ & $\frac{2\rm e^{-x^2}}{\sqrt{\pi}\,erf(x)}+\frac{1}{\sqrt{\pi}\,\rm e^{x^2}x^2erf(x)+x}-\frac{1}{x}$ & $\frac{[1-2xF(x)]^2}{2F^2(x)[F(x)-x]+F(x)}$ & $\frac{1-2x^2}{2x^3+x}$ \\ \\
 $g_{3}(x)$&  $\frac{2\rm e^{-x^2}}{\sqrt{\pi}\,erf(x)}+\frac{1}{\sqrt{\pi}\,\rm e^{x^2}x^2erf(x)+x}-\frac{1}{x}$ &$\frac{1}{F(x)}-2x$& $\frac{1}{x}$ \\ \\
\end{tabular}
\end{center}

\subsection{Even transformation function}

For $u(x)=\rm e^{-x^2/2}\,_1 \mbox{F}_1(\frac{1-2\epsilon}{4};\frac{1}{2};x^2)$ we obtain 
\begin{eqnarray}
\phi_{\varepsilon_1}&\propto& \frac{\,\rm e^{x^2/2}}{\,_1\mbox{F}_1(\frac{1-2\epsilon}{4};\frac{1}{2};x^2)}, \label{Pext2} \\
\phi_{\varepsilon_2}&\propto& u[(\ln u)''-1], \label{Pext3} \\
\phi_{\varepsilon_3}&\propto& \left(1-2\epsilon\right)\,x\,\rm e^{-x^2/2}\,\frac{\,_1\mbox{F}_1(\frac{5-2\epsilon}{4};\frac{3}{2};x^2)}{\,_1\mbox{F}_1(\frac{1-2\epsilon}{4};\frac{1}{2};x^2)}. \label{Pext1}
\end{eqnarray}
These states and their cyclic permutations of $\{\varepsilon_1,\varepsilon_2,\varepsilon_3\}$ lead to the following three
solutions of the PIV equation 
\begin{eqnarray}
g_{1} & = & - 2x + (1 - 2\epsilon)\, x \, \frac{\,_1\mbox{F}_1(\frac{5-2\epsilon}{4};\frac{3}{2};x^2)}{\,_1\mbox{F}_1(\frac{1-2\epsilon}{4};\frac{1}{2};x^2)}, \\
g_{2} & = & - g_1 - 2 x - 2 \left[\frac{x + (2\epsilon - x^2)(g_1+x) + (g_1+x)^3}{x^2-2\epsilon-1-(g_1+x)^2} \right], \\ 
g_{3} & = &  \frac{g_1^2 + 2 x g_1 + 2 \epsilon - 1}{g_1 + 2x},
\end{eqnarray}
which once again belong to the confluent hypergeometric function hierarchy of solutions to the PIV equation with parameters given by the 
expressions found in the previous subsection. For some particular values of $\epsilon$ we get the solutions of the following table:
\\
\begin{center}
\begin{tabular}{c|ccc}
   & $\epsilon=-\frac{1}{2}$ & $\epsilon=\frac{1}{2}$ & $\epsilon=\frac{3}{2}$ \\
    \hline \\
  $g_{1}(x)$ & $0$ & $-2x$ & $\frac{2\left[(1-2x^2)F(x)+x\right]}{2x^2F(x)-1}$ \\
  \\
  $g_{2}(x)$  & $undetermined$ & $0$ & $\frac{\left[(2x^2-1)F(x)-x\right]\left[2\left(x-F(x)\right)F(x)-1\right]}{\left(2xF(x)-1\right)\left[F(x)\left(2x^2F(x)+F(x)-3x\right)+1\right]}$ \\
  \\
  $g_{3}(x)$ & $-\frac{1}{x}$ & $undetermined$ & $\frac{2F(x)}{2xF(x)-1}-\frac{1}{F(x)} $ \\
\end{tabular}
\end{center}

\subsection{2-SUSY}

Recall now that, for the second order SUSY partner Hamiltonians generated from the harmonic oscillator by using as transformation
function $u_1$ and $u_2 = a^- u_1$ with $\epsilon_2=\epsilon_1-1$, there are third order ladder operators $\emph{l}^+$ and 
$\emph{l}^-$ such that the analogue to the number operator factorizes as
\begin{equation}
 \emph{l}^+\emph{l}^-=(H_2-\epsilon_1+1)(H_2-\epsilon_1-1)(H_2-1/2). 
\end{equation}
Therefore, there are three extremal states $\phi_{1}$, $\phi_{2}$ and $\phi_{3}$ with eigenvalues
chosen as $\varepsilon_1=\epsilon_1-1$, $\varepsilon_2=\epsilon_1+1$ and $\varepsilon_3=\frac{1}{2}$,  respectively, which satisfy
\begin{equation}\label{11}
  N\phi_{i}=\emph{l}^+\emph{l}^-\phi_{i}=0, \qquad i=1,2,3.
\end{equation}
Their explicit expressions are given by:
\begin{equation}\label{2.27}
 \phi_{1}\propto \frac{u_1}{W[u_1,u_2]}, \quad  \phi_{2}\propto A^+a^+u_1, \quad  \phi_{3}\propto A^+\chi_0.
\end{equation}
Using the connection formula $u_2(x)=a^-u_1(x)$ with $\epsilon_2=\epsilon_1-1$, 
these states can be expressed in terms of just one transformation function $u_1(x)\equiv u(x)$ with $\epsilon_1=\epsilon$ as follows:
\begin{eqnarray}
\phi_{1}\propto \frac{u}{uu''-(u')^2+u^2} = \frac{1}{u[x^2+1-2\epsilon - (\frac{u'}{u})^2]}, \\
\phi_{2}\propto 2u'-\eta u \\
\phi_{3}\propto \rm e^{-\frac{x^2}{2}}\left[\left(x+\frac{u'}{u}\right)\eta + 2\epsilon - 1\right],
\end{eqnarray}
where
\begin{equation}
 \eta = \frac{2(x+\frac{u'}{u})}{x^2+1-2\epsilon- (\frac{u'}{u})^2}.
\end{equation}

Once again, we can choose any permutation of the indices of $\{\varepsilon_1,\varepsilon_2,\varepsilon_3\}$ in order to 
identify $\phi_{1}$ with any of the three extremal states of the system departing from the choice of 
(\ref{2.27}). Hence we will obtain the following three different solutions of the PIV equation:
\begin{eqnarray}
g_{1} & = &  - x - \alpha + 2\left[\frac{x+\alpha}{x^2+1-2\epsilon-\alpha^2}\right] \\
g_{2} & = &  g_1+\frac{2\alpha^2-2x^2+2(2\epsilon+1)}{\alpha-g_1-x}, \\ 
g_{3} & = & \frac{(x+\alpha)g_1^2+\left[2\epsilon-1+(x+\alpha)^2\right]g_1+(2\epsilon-3)(x+\alpha)}{(x+\alpha)^2+(x+\alpha)g_1+2\epsilon-1}.
\end{eqnarray}
Here we should remember that $\alpha=\frac{u'}{u}$.

Note that $g_{1}$ solves the PIV equation with parameters $a_1=-\epsilon+\frac{5}{2}$ and $b_1=-2\left(\epsilon+\frac{1}{2}\right)^2$
while $g_{2}$ and $g_{3}$ do it for $a_2=-\epsilon-\frac{7}{2}$ and $b_2=-2\left(\epsilon-\frac{3}{2}\right)^2$ 
and $a_3=2(\epsilon-1)$ and $b_3=-8$ respectively. Also, the solutions to PIV depend on our selection of 
the transformation function $u(x)$, which leads once again to two possible options.

\subsection{Odd transformation function}

Taking $u=x\,\rm e^{-x^2/2}\,_1 \mbox{F}_1(\frac{3-2\epsilon}{4};\frac{3}{2};x^2)$ we obtain the following particular solutions
$g_{i}=-x-[\ln\phi_{i}]'$ of the PIV  equation corresponding to different factorization energies $\epsilon$:
\\
\begin{center}
\begin{tabular}{c|ccc}
   & $\epsilon=-\frac{3}{2}$& $\epsilon=\frac{3}{2}$ &$\epsilon=\frac{7}{2} $\\
    \hline\\
 $g_{1}(x)$&$\frac{1+2x^2}{2x^3-x}$&$-\frac{1}{x}-2x$&$\frac{9-48x^4+32x^6-16x^8}{x(-3+2x^2)(3+4x^4)}$\\
  \\
$g_{2}(x)$&$\frac{26x+60x^3-50x^5-56x^7-8x^9}{-1-8x^2+9x^4+20x^6+4x^8}$&$\frac{-4x(2+x^2)}{-5+4x^2+x^4}$&$\frac{4x(243+855x^2-459x^4+168x^6-120x^8+112x^{10}-48x^{12})}{(3+4x^4)(81-162x^2-177x^4+30x^6-28x^8+8x^{10})}$\\
  \\
$g_{2}(x) $&$\frac{4x(3+4x^2-4x^4)}{(-1+2x^2)(3+4x^4)}$&$undetermined$&$\frac{4x(-3+4x^2+4x^4)}{(1+2x^2)(3+4x^4)}$\\
  \end{tabular}
\end{center}

\subsection{Even transformation function}

For $u=\rm e^{-x^2/2}\,_1 \mbox{F}_1(\frac{1-2\epsilon}{4},\frac{1}{2},x^2)$ we get solutions $g_{i}(x)=-x-[\ln\phi_{i}]'$ 
of the PIV equation corresponding to distinct factorization energies $\epsilon$:
\\
\begin{center}
\begin{tabular}{c|ccc}
  &$\epsilon=-\frac{5}{2}$& $\epsilon=-\frac{1}{2}$&$\epsilon=\frac{5}{2}$\\
    \hline \\
 {\scriptsize $g_{1}(x)$}&$\frac{4x(-3+4x^2+4x^4)}{(1+2x^2)(3+4x^4)}$&$0$&$\frac{6x+8x^5}{1-4x^4}$\\
  \\
 {\scriptsize $g_{2}(x)$}&{\scriptsize$\frac{-45+1071x^2+3864x^4+2124x^6+2480x^8+2512x^{10}+768x^{12}+64x^{14}}{-x(3+4x^4)(-15+129x^2+194x^4+76x^6+8x^8)}$}&{\scriptsize$\frac{3+9x^2+2x^4}{-3x-x^3}$}&{\scriptsize$\frac{45-39x^2+14x^4-52x^6-40x^8}{-(45x+107x^3+30x^5-12x^7-8x^9)}$}\\
 \\
 {\scriptsize$g_{3}(x)$}&$\frac{-4x(27+72x^2+16x^8)}{(3+4x^4)(-9+18x^2+12x^4+8x^6)}$&$\frac{4x}{1-2x^2}$&$\frac{4x}{1+2x^2}$\\
 \end{tabular}
\end{center}

\section{Conclusions}

By applying the first-order SUSY QM to the  harmonic oscillator with an infinite potential 
barrier at the origin the supersymmetric partner Hamiltonians, which are isospectral to the initial one, have been 
generated. On the other hand, the second order transformations enlarge the spectral design possibilities for generating
new Hamiltonians with a prescribed spectrum, since now one can either erase a selected level, 
or two consecutive ones. We can also add a new level to the original spectrum almost everywhere, the only restricted 
energies which cannot be produced are the ones corresponding to the even eigenstates of the harmonic oscillator.

When using a first order differential intertwining operator to implement the technique, two choices 
appear for the transformation function $u(x)$ (related to the parity). 

If $u(x)$ is odd there will not be singularities in the generated potential for $x>0$ as long as the factorization 
energy satisfies that $\epsilon\leq\frac{3}{2}=E_0$. Besides, the eigenfunctions of the harmonic oscillator, which 
represent the bound states of the original system, transform into eigenfunctions of the new Hamiltonian. 

On the other hand, if $u(x)$ is even there will not be singularities in the new potentials for $x>0$ as long as the 
factorization energy satisfies that $\epsilon\leq \frac{1}{2}={\cal E}_0$. Moreover, this choice becomes peculiar, 
in the sense that eigenfunctions of the initial Hamiltonian $H_0$ are mapped into NPE of the new 
Hamiltonian $H_1$, while the NPE of $H_0$ transform now into the correct eigenfunctions of $H_1$.

When using a second order differential intertwining operator, four parity combinations for the two 
transformation functions $u_1(x)$ and $u_2(x)$ will appear.

If both $u_1(x)$ and $u_2(x)$ are taken to be odd, no singularities will appear in the transformed potential for 
$x>0$ if the factorization energies are choosen as $\epsilon_2<\epsilon_1\le\frac{3}{2}=E_0$ or 
$E_j=\frac{3+4j}{2}\le\epsilon_2<\epsilon_1\le\frac{3+4(j+1)}{2}=E_{j+1}$. If $u_1(x)$ is odd and $u_2(x)$ is 
even no singularities will appear in the transformed potential for $x>0$ as long as the factorization energies obey that 
${\cal E}_j=\frac{1+4j}{2}\le\epsilon_2<\epsilon_1\le\frac{3+4j}{2}=E_j$. When taking $u_1(x)$ even and $u_2(x)$ odd, 
there will be no extra singularities  in the transformed potential as long as the factorization energies satisfy that 
$\epsilon_2<\epsilon_1\le\frac{1}{2}={\cal E}_0$ or $E_j=\frac{3+4j}{2}\le\epsilon_2<\epsilon_1\le\frac{5+4j}{2} = 
{\cal E}_{j+1}$. Moreover, for these three cases it turns out that the eigenfunctions of the initial Hamiltonian 
transform into eigenfunctions of the new Hamiltonian $H_2$.

When both $u_1(x)$ and $u_2(x)$ are even there will not be singularities in the new potential for $x>0$ if the 
acftorization energies obey that $\epsilon_2<\epsilon_1\le\frac{1}{2}={\cal E}_0$ or 
${\cal E}_j=\frac{1+4j}{2}\le\epsilon_2<\epsilon_1\le\frac{1+4(j+1)}{2} = {\cal E}_{j+1}$. In addition, 
eigenfunctions of the initial Hamiltonian transform into NPE of $H_2$ while the NPE
of $H_0$ transform into the correct eigenfunctions of $H_2$.

Finally, a direct and simple procedure to obtain explicit solutions to the Painlev\'e IV equation was implemented
by using the extremal states for some families of supersymmetric partners of the harmonic oscillator with and infinite 
potential barrier at the origin. Let us note that some rational PIV solutions derived here coincide with several
ones contained in tables 26.1 and 26.2 of \cite{gls02}. A further study of the hierarchies of PIV solutions
which can be generated by applying the SUSY techniques to this truncated harmonic oscillator is still required 
(see however \cite{bf11a,bf11b,be12,bf13}).

\addtolength{\parskip}{\baselineskip}
\textbf{\uppercase{acknowledgments}}

The authors wish to acknowledge the support of CINVESTAV as well as CONACYT through the project 152574 
and the PhD scholarship 243374.

\end{document}